# The Join of the Varieties of R-trivial and L-trivial Monoids via Combinatorics on Words

Manfred Kufleitner[*]     Alexander Lauser[*]

University of Stuttgart, FMI
{kufleitner,lauser}@fmi.uni-stuttgart.de

**Abstract.**   The join of two varieties is the smallest variety containing both. In finite semigroup theory, the varieties of $\mathcal{R}$-trivial and $\mathcal{L}$-trivial monoids are two of the most prominent classes of finite monoids. Their join is known to be decidable due to a result of Almeida and Azevedo. In this paper, we give a new proof for Almeida and Azevedo's effective characterization of the join of $\mathcal{R}$-trivial and $\mathcal{L}$-trivial monoids. This characterization is a single identity of $\omega$-terms using three variables.

**Keywords:**  finite semigroup theory; join of pseudovarieties; Green's relations; combinatorics on words

## 1 Introduction

Green's relations $\mathcal{R}$ and $\mathcal{L}$ are a standard tool in the study of semigroups [5]. In the context of finite monoids, among other results, they have been used to give effective characterizations of language classes such as star-free languages [3, 11] and piecewise testable languages [6, 12]. A deterministic extension of piecewise testable languages yields the class of languages corresponding to $\mathcal{R}$-trivial monoids, and a codeterministic extension corresponds to $\mathcal{L}$-trivial monoids [4, 9].

  Almeida and Azevedo gave an effective characterization for the least variety of finite monoids containing all $\mathcal{R}$-trivial and all $\mathcal{L}$-trivial monoids [2], *i.e.*, for the *join* of the two varieties. Their proof is based on sophisticated algebraic techniques, on Reiterman's Theorem [10], and on a combinatorial result of König [7]. In this paper, we give a new proof of Almeida and Azevedo's Theorem. The current proof was inspired by another proof of the authors [8], which in turn uses ideas of Klíma [6]. The main ingredient is a system of congruences which relies on simple combinatorics on words.

[*]The authors acknowledge the support by the German Research Foundation (DFG) under grant DI 435/5-1.

## 2 Preliminaries

Let $A$ be a finite alphabet. The set of finite words over $A$ is denoted by $A^*$. It is the free monoid over $A$. The *empty word* is 1. The *content* of a word $u = a_1 \cdots a_n$ with $a_i \in A$ is $\alpha(u) = \{a_1, \ldots, a_n\}$, and its *length* is $|u| = n$. The length of the empty word is 0. A word $u$ is a *prefix* (respectively *suffix*) of $v$ if there exists $x \in A^*$ such that $ux = v$ (respectively $xu = v$); if $x \neq 1$, then $u$ is a *proper* prefix.

For more details concerning the algebraic concepts introduced in the remainder of this section, we refer the reader to textbooks such as [1, 4, 9]. *Green's relations* $\mathcal{R}$ and $\mathcal{L}$ are important tools in the study of finite monoids. Let $M$ be a finite monoid. We set $u \mathrel{\mathcal{R}} v$ for $u, v \in M$ if $uM = vM$, and the latter condition is equivalent to the existence of $x, y \in M$ with $u = vx$ and $v = uy$. Symmetrically, $u \mathrel{\mathcal{L}} v$ if $Mu = Mv$. The monoid $M$ is $\mathcal{R}$-*trivial* (respectively $\mathcal{L}$-*trivial*) if $\mathcal{R}$ (respectively $\mathcal{L}$) is the identity relation on $M$. We write $u <_{\mathcal{R}} v$ if $uM \subsetneq vM$, and we write $u <_{\mathcal{L}} v$ if $Mu \subsetneq Mv$.

A *variety* of finite monoids is a class of monoids closed under finite direct products, submonoids, and quotients. A variety of finite monoids is often called a *pseudovariety* in order to distinguish from varieties in Birkhoff's sense. Since we do not need this distinction in the current paper, whenever we use the term *variety* we mean a variety of finite monoids. The *join* $\mathbf{V}_1 \vee \mathbf{V}_2$ of two varieties $\mathbf{V}_1$ and $\mathbf{V}_2$ is the smallest variety containing $\mathbf{V}_1 \cup \mathbf{V}_2$. A monoid $M$ is in $\mathbf{V}_1 \vee \mathbf{V}_2$ if and only if there exist $M_1 \in \mathbf{V}_1$ and $M_2 \in \mathbf{V}_2$ such that $M$ is a quotient of a submonoid of $M_1 \times M_2$. If $M$ is a finite monoid, then there exists an integer $\omega_M \geqslant 1$ such that, for all $u \in M$, the element $u^{\omega_M}$ is idempotent. Moreover, the element $u^{\omega_M}$ is the unique idempotent generated by $u$. Usually, the monoid $M$ is clear from the context and thus, we simply write $\omega$ instead of $\omega_M$. This leads to the following definition. An $\omega$-*term* over a finite alphabet $X$ is either a word in $X^*$, or of the form $t^\omega$ for some $\omega$-term $t$, or the concatenation $t_1 t_2$ of two $\omega$-terms $t_1, t_2$. A homomorphism $\varphi : X^* \to M$ to a finite monoid $M$ uniquely extends to $\omega$-terms over $X$ by setting $\varphi(t^\omega) = \varphi(t)^{\omega_M}$. Let $u, v$ be two $\omega$-terms over $X$. A finite monoid $M$ *satisfies* the identity $u = v$ if $\varphi(u) = \varphi(v)$ for all homomorphisms $\varphi : X^* \to M$. The class of finite monoids satisfying the identity $u = v$ is denoted by $[\![ u = v ]\!]$. For all $\omega$-terms $u, v$, the class $[\![ u = v ]\!]$ forms a variety. We need the following three varieties in this paper:

$$\begin{aligned}
\mathbf{R} &= [\![ (xy)^\omega x = (xy)^\omega ]\!], \\
\mathbf{L} &= [\![ x(zx)^\omega = (zx)^\omega ]\!], \\
\mathbf{W} &= [\![ (xy)^\omega x(zx)^\omega = (xy)^\omega (zx)^\omega ]\!].
\end{aligned}$$

A monoid is in $\mathbf{R}$ if and only if it is $\mathcal{R}$-trivial. Symmetrically, a monoid is in $\mathbf{L}$ if and only if it is $\mathcal{L}$-trivial. The aim of this paper is to give a new proof of Almeida and Azevedo's result $\mathbf{R} \vee \mathbf{L} = \mathbf{W}$. The inclusion $\mathbf{R} \vee \mathbf{L} \subseteq \mathbf{W}$ is trivial since $\mathbf{R} \cup \mathbf{L} \subseteq \mathbf{W}$ and $\mathbf{W}$ is a variety.

## 3 Congruences

In this section, we introduce the main combinatorial tool for our proof. It is a family of congruences $\equiv_n$ on $A^*$ for some finite alphabet $A$ such that $A^*/\equiv_n \in \mathbf{R} \vee \mathbf{L}$ for all integers $n \geqslant 0$, see Lemma 2 below. As a first step towards the definition of $\equiv_n$ we need to introduce an asymmetric, weaker congruence $\equiv_n^{\mathcal{R}}$.

Let $u, v \in A^*$. We let $u \equiv_0^{\mathcal{R}} v$ if $\alpha(u) = \alpha(v)$. For $n \geqslant 0$, we let $u \equiv_{n+1}^{\mathcal{R}} v$ if the following conditions hold:

1. $\alpha(u) = \alpha(v)$,
2. for all factorizations $u = u_1 a u_2$ and $v = v_1 a v_2$ with $a \in A \setminus \big(\alpha(u_1) \cup \alpha(v_1)\big)$ we have $u_1 \equiv_n^{\mathcal{R}} v_1$ and $u_2 \equiv_n^{\mathcal{R}} v_2$, and
3. for all factorizations $u = u_1 a u_2$ and $v = v_1 a v_2$ with $a \in A \setminus \big(\alpha(u_2) \cup \alpha(v_2)\big)$ we have $u_1 \equiv_n^{\mathcal{R}} v_1$.

By a straightforward verification we see that $\equiv_n^{\mathcal{R}}$ is an equivalence relation. The factorization $u_1 a u_2$ with $a \in A \setminus \alpha(u_1)$ is unique. Therefore, induction on $n$ shows that the index of $\equiv_n^{\mathcal{R}}$ is finite. If $u \equiv_{n+1}^{\mathcal{R}} v$, then $u \equiv_n^{\mathcal{R}} v$. Moreover, if $u \equiv_n^{\mathcal{R}} v$ and $a \in A$, then $au \equiv_n^{\mathcal{R}} av$ and $ua \equiv_n^{\mathcal{R}} va$. Therefore, the relation $\equiv_n^{\mathcal{R}}$ is a finite index congruence on $A^*$.

**Lemma 1** *For every finite alphabet $A$ and every integer $n \geqslant 0$ we have $A^*/{\equiv_n^{\mathcal{R}}} \in \mathbf{R}$.*

*Proof.* It suffices to show $(xy)^{n+1}x \equiv_n^{\mathcal{R}} (xy)^{n+1}$ for all words $x, y \in A^*$. We note that for $y = 1$ this yields $x^{n+2} \equiv_n^{\mathcal{R}} x^{n+1}$. The proof is by induction on $n$. For $n = 0$, the claim is true since $\alpha(xyx) = \alpha(xy)$. Let now $n > 0$. As before, $\alpha\big((xy)^{n+1}x\big) = \alpha\big((xy)^{n+1}\big)$. Suppose $(xy)^{n+1}x = u_1 a u_2$ and $(xy)^{n+1} = v_1 a v_2$ for $a \in A \setminus \big(\alpha(u_1) \cup \alpha(v_1)\big)$. Then $u_1 = v_1$ and both are proper prefixes of $xy$. Thus $u_2 = p(xy)^n x$ and $v_2 = p(xy)^n$ for some $p \in A^*$. By induction $(xy)^n x \equiv_{n-1}^{\mathcal{R}} (xy)^n$ and hence, $u_2 \equiv_n^{\mathcal{R}} v_2$.

Suppose now $(xy)^{n+1}x = u_1 a u_2$ and $(xy)^{n+1} = v_1 a v_2$ for $a \in A \setminus \big(\alpha(u_2) \cup \alpha(v_2)\big)$. Then $av_2$ is a suffix of $xy$ and $au_2$ is a suffix of $yx$. We can therefore write $v_1 = (xy)^n p'$ for some prefix $p'$ of $xy$. Similarly, $u_1 = (xy)^k p$ for some $k \in \{n, n+1\}$ and some prefix $p$ of $xy$, i.e., we have $pq = xy$ for some $q \in A^*$. By induction, we have $(xy)^{n+1} \equiv_{n-1}^{\mathcal{R}} (xy)^n$ and thus $(xy)^{n+1}p \equiv_{n-1}^{\mathcal{R}} (xy)^n p$. We can therefore assume $k = n$. Without loss of generality, let $|p| \leqslant |p'|$, i.e., $p' = ps$ for some $s \in A^*$. It follows

$$u_1 = (pq)^n p \quad \text{and} \quad v_1 = (pq)^n ps.$$

Since $p' = ps$ is a prefix of $xy = pq$, the word $s$ is a prefix of $q$. In particular, there exists $t \in A^*$ such that $qp = st$. This yields

$$u_1 = p(st)^n \quad \text{and} \quad v_1 = p(st)^n s.$$

By induction, $(st)^n \equiv_{n-1}^{\mathcal{R}} (st)^n s$ and thus $u_1 \equiv_{n-1}^{\mathcal{R}} v_1$. This shows $(xy)^{n+1}x \equiv_n^{\mathcal{R}} (xy)^{n+1}$ which concludes the proof. □

There is a left-right symmetric congruence $\equiv_n^{\mathcal{L}}$ on $A^*$. It can be defined by setting $u \equiv_n^{\mathcal{L}} v$ if and only if $u^\rho \equiv_n^{\mathcal{R}} v^\rho$. Here, $u^\rho = a_n \cdots a_1$ is the *reversal* of the word $u = a_1 \cdots a_n$ with $a_i \in A$. It satisfies $A^*/{\equiv_n^{\mathcal{L}}} \in \mathbf{L}$ for every $n \geqslant 0$. We define $u \equiv_n v$ if and only if both $u \equiv_n^{\mathcal{R}} v$ and $u \equiv_n^{\mathcal{L}} v$. The following lemma puts together some properties of the finite index congruence $\equiv_n$.

**Lemma 2** *For every finite alphabet $A$ and every integer $n \geqslant 0$ the following properties hold:*

1. $A^*/{\equiv_n} \in \mathbf{R} \vee \mathbf{L}$.
2. *If $u_1 a u_2 \equiv_{n+1} v_1 a v_2$ for $a \in A \setminus \big(\alpha(u_1) \cup \alpha(v_1)\big)$, then $u_1 \equiv_n^{\mathcal{R}} v_1$ and $u_2 \equiv_n v_2$.*
3. *If $u_1 a u_2 \equiv_{n+1} v_1 a v_2$ for $a \in A \setminus \big(\alpha(u_2) \cup \alpha(v_2)\big)$, then $u_1 \equiv_n v_1$ and $u_2 \equiv_n^{\mathcal{L}} v_2$.*

*Proof.* "*1*": We have $A^*/{\equiv_n} \in \mathbf{R} \vee \mathbf{L}$ since it is a submonoid of $(A^*/{\equiv_n^{\mathcal{R}}}) \times (A^*/{\equiv_n^{\mathcal{L}}})$, and $A^*/{\equiv_n^{\mathcal{R}}} \in \mathbf{R}$ and $A^*/{\equiv_n^{\mathcal{L}}} \in \mathbf{L}$ by Lemma 1 and its left-right dual. The properties "*2*" and "*3*" trivially follow from the definition of $\equiv_n$. □

# 4 An Equation for the Join

The goal of this section is to prove $\mathbf{W} \subseteq \mathbf{R} \vee \mathbf{L}$. By Lemma 2 it suffices to show that for every $A$-generated monoid $M \in \mathbf{W}$ there exists an integer $n \geqslant 0$ such that $M$ is a quotient of $A^*/{\equiv_n}$. The outline of the proof is as follows. First, in Lemma 3, we give a substitution rule valid in $\mathbf{W}$. Then, in Lemma 5, we show that $\equiv_n$-equivalence allows a factorization satisfying the premise for applying this substitution rule; this relies on a property of $\mathbf{W}$ shown in Lemma 4. Finally, in Theorem 6, all the ingredients are put together.

**Lemma 3** *Let $M \in \mathbf{W}$ and let $u, v, x \in M$. If $u \mathcal{R} ux$ and $v \mathcal{L} xv$, then $uxv = uv$.*

*Proof.* Since $u \mathcal{R} ux$ and $v \mathcal{L} xv$, there exist $y, z \in M$ with $u = uxy$ and $v = zxv$. In particular, we have $u = u(xy)^\omega$ and $v = (zx)^\omega v$. By $M \in \mathbf{W}$ we conclude $uxv = u(xy)^\omega x(zx)^\omega v = u(xy)^\omega (zx)^\omega v = uv$. □

We will apply the previous lemma as follows. Let $M \in \mathbf{W}$ and $u, v, s, t \in M$ such that $u \mathcal{R} us \mathcal{R} ut$ and $v \mathcal{L} sv \mathcal{L} tv$. Then $usv = utv$ since $usv = uv$ and $utv = uv$ by Lemma 3. The $\mathcal{R}$-equivalences and $\mathcal{L}$-equivalences for being able to apply this substitution rule are established in Lemma 5. Before, we give a simple property of $\mathbf{W}$. It is the link between Green's relations and the congruence $\equiv_n$.

**Lemma 4** *Let $M \in \mathbf{W}$ and let $u, v, a \in M$. If $u \mathcal{R} v \mathcal{R} va$, then $u \mathcal{R} ua$. If $u \mathcal{L} v \mathcal{L} av$, then $u \mathcal{L} au$.*

*Proof.* Since $u \mathcal{R} v$ and $u \mathcal{R} va$, there exist $x, y \in M$ with $v = ux$ and $u = vay$. Now, $u = uxay = u(xay)^{2\omega+1} = u(xay)^\omega x(ayx)^\omega ay = u(xay)^\omega (ayx)^\omega ay = u(ayx)^\omega ay \in uaM$ where the fourth equality uses $M \in \mathbf{W}$. This shows $uM \subseteq uaM$ and thus $u \mathcal{R} ua$. The second implication is left-right symmetric. □

The intuitive interpretation of the algebraic statement in Lemma 4 is the following: For $M \in \mathbf{W}$ it only depends on the element $a$ and the $\mathcal{R}$-class of $u$ whether $u \mathcal{R} ua$ or not (but not on the element $u$ itself). The statement for $\mathcal{L}$-classes is analogous.

**Lemma 5** *Let $M \in \mathbf{W}$ and let $\varphi : A^* \to M$ be a homomorphism. If $u \equiv_n v$ for $n \geqslant 2|M|$, then there exist factorizations $u = a_1 s_1 \cdots a_{\ell-1} s_{\ell-1} a_\ell$ and $v = a_1 t_1 \cdots a_{\ell-1} t_{\ell-1} a_\ell$ with $a_i \in A$ and $s_i, t_i \in A^*$ and with $\ell \leqslant 2|M|$ such that for all $i \in \{1, \ldots, \ell-1\}$ we have:*

$$\varphi(a_1 s_1 \cdots a_{i-1} s_{i-1} a_i) \mathcal{R} \varphi(a_1 s_1 \cdots a_i s_i) \mathcal{R} \varphi(a_1 s_1 \cdots a_{i-1} s_{i-1} a_i t_i),$$
$$\varphi(a_{i+1} t_{i+1} \cdots a_{\ell-1} t_{\ell-1} a_\ell) \mathcal{L} \varphi(t_i a_{i+1} \cdots t_{\ell-1} a_\ell) \mathcal{L} \varphi(s_i a_{i+1} t_{i+1} \cdots a_{\ell-1} t_{\ell-1} a_\ell).$$

*Proof.* To simplify notation, for some relation $\mathcal{G}$ on $M$ we write $u \mathcal{G} v$ for words $u, v \in A^*$ if $\varphi(u) \mathcal{G} \varphi(v)$. Consider the $\mathcal{R}$-factorization of $u$, i.e., let $u = b_1 u_1 \cdots b_k u_k$ with $b_i \in A$ such that

$$b_1 u_1 \cdots b_i \; \mathcal{R} \; b_1 u_1 \cdots b_i u_i \qquad \text{for all } i \in \{1, \ldots, k\},$$
$$b_1 u_1 \cdots b_i u_i >_\mathcal{R} b_1 u_1 \cdots b_i u_i b_{i+1} \qquad \text{for all } i \in \{1, \ldots, k-1\}.$$

Similarly, let $v = v_1 c_1 \cdots v_{k'} c_{k'}$ be the $\mathcal{L}$-factorization of $v$, i.e., we have $c_i \in A$ and

$$c_i \cdots v_{k'} c_{k'} \; \mathcal{L} \; v_i c_i \cdots v_{k'} c_{k'} \qquad \text{for all } i \in \{1, \ldots, k'\},$$
$$v_i c_i \cdots v_{k'} c_{k'} >_\mathcal{L} c_{i-1} v_i c_i \cdots v_{k'} c_{k'} \qquad \text{for all } i \in \{2, \ldots, k'\}.$$

We have $k, k' \leqslant |M|$ because neither the number of $\mathcal{R}$-classes nor the number of $\mathcal{L}$-classes can exceed $|M|$. By Lemma 4, we have $b_i \notin \alpha(u_{i-1})$ for all $i \in \{2, \ldots, k\}$ and $c_i \notin \alpha(v_{i+1})$ for all $i \in \{1, \ldots, k'-1\}$. We use these properties to convert the $\mathcal{R}$-factorization of $u$ to $v$ and to convert the $\mathcal{L}$-factorization of $v$ to $u$: Let $v = b_1 v_1' \cdots b_k v_k'$ such that $b_i \notin \alpha(v_{i-1}')$, and let $u = u_1' c_1 \cdots u_{k'}' c_{k'}$ with $c_i \notin \alpha(u_{i+1}')$. These factorizations exist because $u \equiv_n v$; in particular, by Lemma 2,

$$u_i b_{i+1} u_{i+1} \cdots b_k u_k \equiv_{n-i} v_i' b_{i+1} v_{i+1}' \cdots b_k v_k'$$
$$v_1 c_1 \cdots v_{j-1} c_{j-1} v_j \equiv_{n-k'-1+j} u_1' c_1 \cdots u_{j-1}' c_{j-1} u_j'$$

for all $i \in \{1, \ldots k\}$ and $j \in \{1, \ldots, k'\}$. Moreover, we see $\alpha(u_i) = \alpha(v_i')$ and $\alpha(v_j) = \alpha(u_j')$.

We now show that the relative positions of the $b_i$'s and $c_j$'s in the above factorizations are the same in $u$ and $v$. Let $p$ be the position of $b_i$ in the $\mathcal{R}$-factorization of $u$ and let $q$ be the position of $c_j$ in the above factorization of $u$. Similarly, let $p'$ be the position of $b_i$ in $v$ and let $q'$ be the position of $c_j$ in $v$. First, suppose $p < q$. Let

$$u = b_1 u_1 \cdots b_{i-1} u_{i-1} b_i \, u' \, c_j u_{j+1}' c_{j+1} \cdots u_{k'}' c_{k'}.$$

By an $i$-fold application of property "2" in Lemma 2 with $a \in \{b_1, \ldots, b_i\}$ (which is possible for $u$) we obtain $v = b_1 v_1' \cdots b_{i-1} v_{i-1}' b_i z$ with $z \equiv_{n-i} u' c_j u_{j+1}' c_{j+1} \cdots u_{k'}' c_{k'}$. By a $(k'+1-j)$-fold application of property "3" in Lemma 2 with $a \in \{c_{k'}, \ldots, c_j\}$ (which is possible for the word $u' c_j u_{j+1}' c_{j+1} \cdots u_{k'}' c_{k'}$) we obtain $z = v' c_j v_{j+1} c_{j+1} \cdots v_{k'} c_{k'}$. Thus

$$v = b_1 v_1' \cdots b_{i-1} v_{i-1}' b_i \, v' \, c_j v_{j+1} c_{j+1} \cdots v_{k'} c_{k'}$$

showing that $p' < q'$. Symmetrically, one shows that $p' < q'$ implies $p < q$. We conclude $p < q$ if and only if $p' < q'$. Similarly, we have $p = q$ if and only if $p' = q'$. It follows that the relative order of the $b_i$'s and $c_j$'s in $u$ and $v$ is the same. By factoring $u$ and $v$ at all $b_i$'s and $c_j$'s, we obtain $u = a_1 s_1 \cdots a_{\ell-1} s_{\ell-1} a_\ell$ and $v = a_1 t_1 \cdots a_{\ell-1} t_{\ell-1} a_\ell$ with $a_i \in A$ and $\ell \leqslant k + k' \leqslant 2|M|$.

We have $a_1 s_1 \cdots a_{i-1} s_{i-1} a_i \; \mathcal{R} \; a_1 s_1 \cdots a_{i-1} s_{i-1} a_i s_i$ because $u = a_1 s_1 \cdots a_{\ell-1} s_{\ell-1} a_\ell$ is a refinement of the $\mathcal{R}$-factorization. Note that we cannot assume $\alpha(s_i) = \alpha(t_i)$. But each $t_i$ is a factor of some $v_j'$, and at the same time $s_i$ is a factor of $u_j$. More precisely, there exists $m \leqslant i$ such that

$$b_1 v_1' \cdots b_{j-1} v_{j-1}' b_j = a_1 t_1 \cdots a_{m-1} t_{m-1} a_m \quad \text{and} \quad t_m a_{m+1} \cdots t_{i-1} a_i t_i \; \text{is a prefix of} \; v_j'.$$

Furthermore, $s_m a_{m+1} \cdots s_{i-1} a_i s_i$ is a prefix of $u_j$. Now, $\alpha(t_i) \subseteq \alpha(v_j') = \alpha(u_j)$ and, by Lemma 4, for all words $z$ with $\alpha(z) \subseteq \alpha(u_j)$ we have $a_1 s_1 \cdots a_{i-1} s_{i-1} a_i \; \mathcal{R} \; a_1 s_1 \cdots a_{i-1} s_{i-1} a_i z$. Symmetrically we see $a_{i+1} t_{i+1} \cdots a_{\ell-1} t_{\ell-1} a_\ell \; \mathcal{L} \; t_i a_{i+1} \cdots t_{\ell-1} a_\ell \; \mathcal{L} \; s_i a_{i+1} t_{i+1} \cdots a_{\ell-1} t_{\ell-1} a_\ell$. □

**Theorem 6 (Almeida/Azevedo, 1989 [2])**

$$\mathbf{R} \vee \mathbf{L} = [\![ (xy)^\omega x (zx)^\omega = (xy)^\omega (zx)^\omega ]\!]$$

*Proof.* The inclusion $\mathbf{R} \vee \mathbf{L} \subseteq \mathbf{W}$ is trivial since $\mathbf{R} \cup \mathbf{L} \subseteq \mathbf{W}$ and $\mathbf{W}$ is a variety of finite monoids. Let $M \in \mathbf{W}$ be generated by $A$, and let $\varphi : A^* \to M$ be the homomorphism induced by $A \subseteq M$. Let $n = 2\,|M|$ and suppose $u \equiv_n v$. Let $u = a_1 s_1 \cdots a_{\ell-1} s_{\ell-1} a_\ell$ and $v = a_1 t_1 \cdots a_{\ell-1} t_{\ell-1} a_\ell$ be the factorizations from Lemma 5. Applying Lemma 3 repeatedly, we get

$$\begin{aligned}
\varphi(v) &= \varphi(a_1 t_1 a_2 t_2 \cdots a_{\ell-2} t_{\ell-2} a_{\ell-1} t_{\ell-1} a_\ell) \\
&= \varphi(a_1 s_1 a_2 t_2 \cdots a_{\ell-2} t_{\ell-2} a_{\ell-1} t_{\ell-1} a_\ell) \\
&= \varphi(a_1 s_1 a_2 s_2 \cdots a_{\ell-2} t_{\ell-2} a_{\ell-1} t_{\ell-1} a_\ell) \\
&\quad\vdots \\
&= \varphi(a_1 s_1 a_2 s_2 \cdots a_{\ell-2} s_{\ell-2} a_{\ell-1} t_{\ell-1} a_\ell) \\
&= \varphi(a_1 s_1 a_2 s_2 \cdots a_{\ell-2} s_{\ell-2} a_{\ell-1} s_{\ell-1} a_\ell) = \varphi(u).
\end{aligned}$$

Note that the substitution rules $t_i \to s_i$ are $\varphi$-invariant only when applied from left to right. This shows that $M$ is a quotient of $A^*/\equiv_n$, and the latter is in $\mathbf{R} \vee \mathbf{L}$ by Lemma 2. Thus $M \in \mathbf{R} \vee \mathbf{L}$. □